\documentclass[draft]{agujournal2019}
\usepackage{url}
\usepackage{lineno}
\usepackage[inline]{trackchanges} 
\usepackage{soul}
\usepackage{pdfpages}
\newcommand{\CO}{CO$_2$ }

\newcommand{\Ls}{L$_s$}

\draftfalse

\journalname{GRL}

\begin{document}

\title{A Reappraisal of  Subtropical Subsurface Water Ice Stability on Mars}

\authors{L.Lange\affil{1}, F.Forget\affil{1}, M.Vincendon\affil{2}, A.Spiga\affil{1}, E.Vos\affil{1,3}, O.Aharonson\affil{3},
E.Millour\affil{1},A.Bierjon\affil{1}, R.Vandemeulebrouck\affil{1}}

\affiliation{1}{Laboratoire de Météorologie Dynamique, Institut Pierre-Simon Laplace (LMD/IPSL), Sorbonne Université, Centre National de la Recherche Scientifique (CNRS), École Polytechnique, École Normale Supérieure (ENS), Paris, France}
\affiliation{2}{Institut d’Astrophysique Spatiale, Université Paris-Saclay, CNRS, Orsay, France}
\affiliation{3}{Department of Earth and Planetary Sciences, Weizmann Institute of Science, Rehovot, Israel}

\correspondingauthor{Lucas Lange}{lucas.lange@lmd.ipsl.fr}

\begin{keypoints}
\item We use a new model of steep slope microclimates to explore the stability of subsurface water ice on Mars at latitudes lower than 30\textdegree;

\item Our model shows that warm plains and large-scale atmospheric dynamics heat these slopes, preventing ice from being stable; 

\item Subsurface ice is predicted to be present down to 30\textdegree~of latitude, possibly down to 25\textdegree~but for sparse slopes with favorable conditions.
\end{keypoints}

\begin{abstract}

Massive reservoirs of subsurface water ice in equilibrium with  atmospheric water vapor are found poleward of 45\textdegree~latitude on Mars.  The absence of CO$_2$ frost on steep pole-facing slopes and simulations of atmospheric-soil water exchanges suggested that water ice could be stable underneath these slopes down to 25\textdegree~latitude. We revisit these arguments with a new slope microclimate model. Our model shows that below 30\textdegree~latitude, slopes are warmer than previously estimated as the air above is heated by warm surrounding plains. This additional heat prevents the formation of  surface \CO frost and subsurface water ice for most slopes. Our model suggests the presence of subsurface water ice beneath pole-facing slopes down to 30\textdegree~latitude, and possibly  25$^\circ$ latitude on sparse steep dusty slopes. While  unstable ice deposits might be present,  our results suggest that water ice is rarer than previously thought in the $\pm$30\textdegree~latitude range considered for human exploration.

 \end{abstract}

\section*{Plain Language Summary}
The presence of water ice near the equator is a key issue for future human exploration of Mars. In the current climate, this ice cannot exist near the equator but could be stable  at accessible depths below pole-facing slopes down to latitudes of 25\textdegree, i.e., close enough to the equator for a crewed mission. Here, we study the possible presence of this subsurface ice  with a new model that simulates the microclimates associated with slopes on Mars. Our results show that, contrary to the arguments put forward in the literature, the slopes close to the equator (20\textdegree-30\textdegree) may in fact be too warm to allow subsurface water ice to be stable, and that the observations that suggested the presence of ice under these slopes can be explained otherwise by our model.  Thus, the  widespread presence of water ice under these slopes at  subtropical latitudes is not demonstrated. However, our model cannot rule out the presence of ancient ice reservoirs, that would be slowly sublimating today.

\section{Introduction}

The Martian global water inventory is distributed in five reservoirs: the atmosphere, the surface ice, adsorbed water, hydrated minerals, and subsurface  ice. The global average water content of the atmosphere, controlled by the sublimation of the northern polar cap, is about 10~pr-$\mu$m (expressed as precipitable microns of total column abundance)
\cite{Smith2002}, while the perennial polar deposits represent 2/3 of the global exchangeable water inventory \cite<nearly 3.2-4.7$\times$10$^6$~km$^3$ of water for both caps,>{Montmessin2017}. Depending on the regolith properties, the adsorbed water content can be up to 100~pr-$\mu$m \cite{Montmessin2017}.  
The  Mars Odyssey Neutron Spectrometer \cite{Boynton2004} and Fine-Resolution Epithermal Neutron Detector  \cite{Mitrofanov2018} revealed a significant amount of water in the shallow subsurface ($\leq$~1~m~depth)  at high latitudes \cite{Boynton2002, Feldman2002science,  Malakhov2020}. The total amount of water in this reservoir is poorly constrained because its bottom depth is not  known. However, computations suggested that the upper meter of high-latitude regolith could contain nearly 10$^4$ more water than the atmosphere \cite{Montmessin2017}. Finally, hydrated minerals could contain up to 5 times the water content of all other water reservoirs combined \cite{Wernicke2021}.

Subsurface water ice (hereinafter referred to as subsurface ice) is of significant interest for the understanding and exploration of Mars. First, it affects the seasonal condensation and sublimation of polar caps, and thus directly impacts the \CO cycle \cite{Haberle2008}: because ground ice has a large thermal inertia, it stores heat during summer and releases it during winter, reducing the condensation rate of \CO ice.   Second, this ice can record the history of volatile transport across water reservoirs \cite{Vos2022}.  Finally, subsurface ice represents a major exploitable resource for future crewed exploration, as part of a strategy to rely on in-situ resources \cite{Morgan2021}. Thermal requirements for future crewed missions limit the possible landing site to latitudes lower than 30\textdegree~\cite{GRANT2018,Morgan2021}. The MONS spectrometer has not observed ice enrichment in the first meter at such low latitudes \cite{DIEZ2008}, even if a recent impact revealed shallow subsurface ice down to 35\textdegree N latitude, which is the closest detection to the equator to date \cite{Dundas2022}.   The existence and characterization of  ice at low latitudes are therefore a significant challenge today \cite{Bramson2021,Putzig2023}.   

As favorable landing conditions are frequently located at latitudes lower than 30\textdegree, we focus here on an assessment of subsurface ice stability  within a 20\textdegree~to 30\textdegree~latitude band. We will notably assess the stability of ice between 25\textdegree~and 30\textdegree, a latitude range that will be referred to as "\textit{subtropical latitudes}" in the following part of the paper. At such  latitudes, subsurface ice is not expected to be stable on flat terrains \cite{Schorghofer2005} but could be stable on pole-facing slopes which have cold microclimates. Two types of studies have suggested the presence of ice at subtropical latitudes under pole-facing terrains:

\begin{enumerate}

    \item  \citeA{Vincendon2010} studied the stability of \CO ice on  pole-facing slopes at mid and subtropical latitudes in the Southern hemisphere. Seasonal \CO ice was not observed on slopes for latitudes lower than 34\textdegree~while 1D thermal modeling indicated that ice should be present. They showed that the most likely explanation for the absence of frost at these latitudes lower than 34\textdegree, and the narrow distribution of \CO ice observations between 45° and 34\textdegree, was the presence of a latitude-dependent high thermal inertia material (most likely buried water ice with a latitude-dependent depth)  under these slopes,  which released heat during the winter and made the \CO ice unstable.

    \item Numerical models by \citeA{Aharonson2006} and \citeA{Mellon2022}  of subsurface  ice stability showed that ground ice could be stable with respect to diffusion down to 25\textdegree~of latitude  on steep pole-facing slopes.

\end{enumerate}

Here, we show that  subsurface ice is probably not stable on pole-facing slopes at latitudes lower than 30\textdegree, except in sparse locations with very favorable conditions and down to 25\textdegree~latitude only (high slope angle, low thermal inertia, and high albedo).  The model that is used to simulate slope microclimates and the subsurface ice stability is presented in section \ref{sec:Method}. Using this new model,  we show in section \ref{ssec:Vincendonstudy} that the absence of \CO ice on low latitude slopes can be explained without requiring the presence of subsurface  ice. We then apply our subsurface ice model to compute the theoretical water ice stability in section \ref{ssec:Ahronsonstudy} and show that previous studies  may have overestimated the latitudinal extent of stable subsurface ice. The possible presence of subsurface ice in sparse favorable locations is discussed in section \ref{sec:Discussion}. Conclusions are drawn in section \ref{sec:conlusion}.

\section{Methods and Model \label{sec:Method}}

\subsection{Mars Planetary Climate Model \label{ssec:PCM}}
The current study uses the Mars Planetary Climate Model (PCM) version 6 \cite{Forget1999, Forget2022}. We added  a sub-grid slope parameterization to simulate slope microclimates. This parametrization is detailed and compared to observations in a companion paper  \cite{Lange2023Model} that is summarized in Text S1.  In short: for each PCM mesh, we decompose the cell as a distribution of sloped terrains (defined by characteristic slopes) and a flat terrain. On each sub-grid terrain, we compute the radiative transfer following \citeA{Spiga2011}, turbulent exchanges \cite{Forget1999}, and the condensation of volatiles \cite{Forget1998snow, Navarro2014}. The portion of the atmosphere above the ground within the cell is in equilibrium with a weighted average of these surface microclimates. 
Surface properties (albedo, emissivity, thermal inertia) are set to the observations from the Thermal Emission Spectrometer  \cite<TES, >{Christensen2001, PUTZIG2007}.  A nominal dust opacity scenario is used \cite{MONTABONE2015}.

\subsection{Subsurface Ice Model \label{ssec:SSI}}
The model used  to compute the ice table depth at equilibrium follows the approach of \citeA{ Schorghofer2005, Aharonson2006, Mellon2022}:  subsurface ice is stable at a depth \textit{z}  if:

\begin{equation}
    \overline{\left(\frac{p_{\rm{vap,surf}}}{T_{\rm{surf}}}\right)} \geq \overline{\left(\frac{p_{\rm{sv}}(T_{\rm{soil}}(z))}{T_{\rm{soil}}(z)}\right)}
    \label{eq:stabilityssi}
\end{equation}

\noindent where overbars indicate time-averages over a complete martian year,  $p_{\rm{vap, surf}}$~(Pa) is the vapor pressure at the surface, $T_{\rm{surf}}$~(K) is the surface temperature, $p_{\rm{sv, soil}} $~(Pa) is the saturation vapor pressure which is a function of the soil temperature $T_{\rm{soil}}$~(K) \cite{Murphy2005}. $p_{\rm{vap, surf}}$ is computed by the PCM, whose water cycle has been fully validated \cite{Navarro2014, Naar2021} as well as near-surface vapor content through comparison with Phoenix measurements \cite{Fischer2019}. In this model, we also include the effect of surface water frost that stabilizes the ice table \cite{Hagedorn2007, McKay2009, Williams2015}.  When ice is stable at depth $z_{ice}$, we set the thermal inertia of this layer and those below to  1,600~J~m$^{-2}$~K$^{-1}$~s$^{-1/2}$, a mid-value between completely pore-filled ice and massive pure ice \cite{Schorghofer2005, Siegler2012}.   This model is run for tens of years until the ice table depth has reached an equilibrium.

\section{Results}
\subsection{\CO ice stability on subtropical slopes \label{ssec:Vincendonstudy}}

\begin{figure}[ht!]
    \centering
    \includegraphics[width = \textwidth]{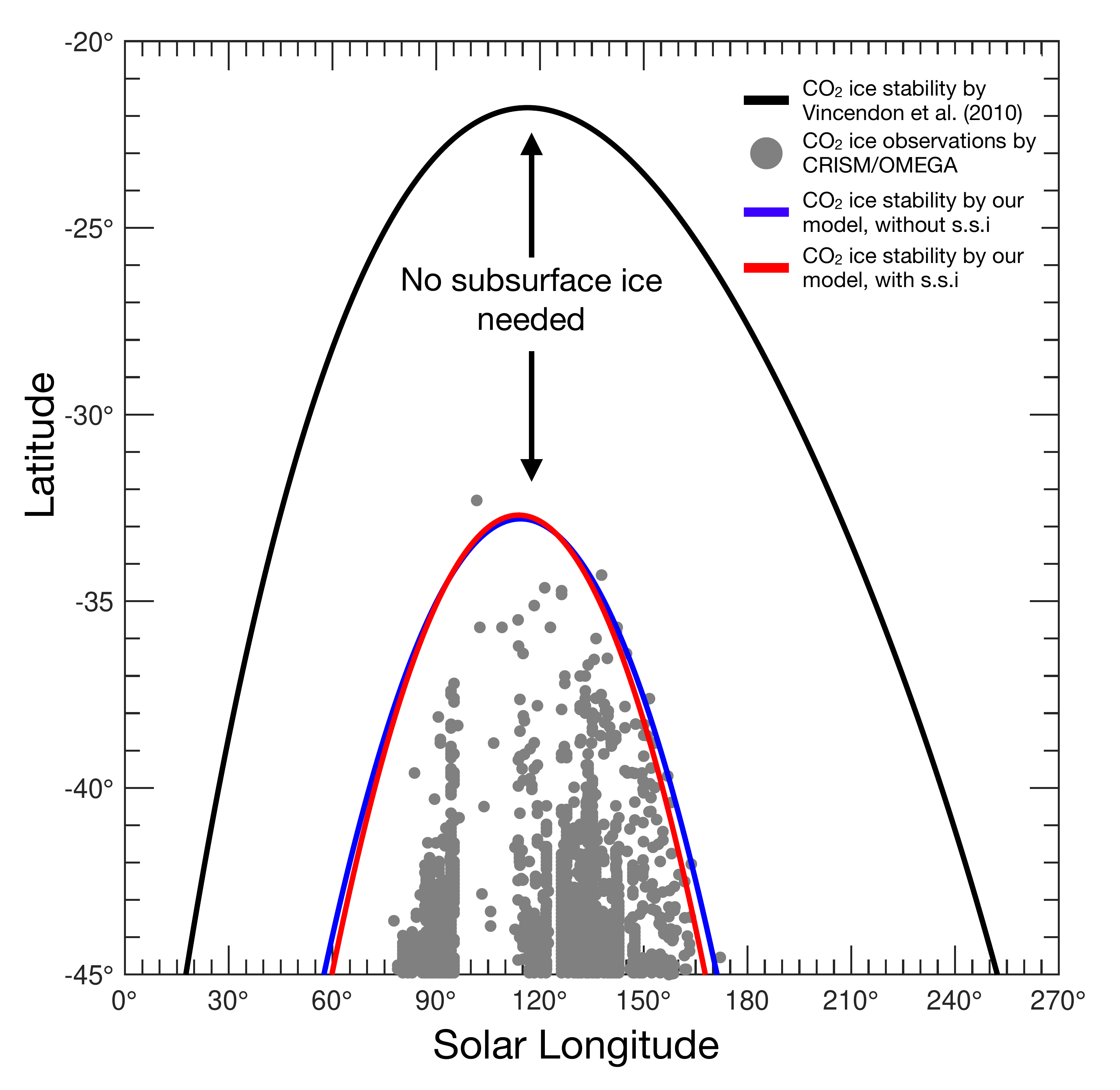}
    \caption{\CO ice stability on 30\textdegree~pole-facing slopes in the Southern hemisphere as a function of the solar longitude (\Ls, the Mars-Sun angle, measured from the Northern Hemisphere spring equinox where \Ls=0\textdegree). 
    The black curve corresponds to the stability predicted by the 1D model of \citeA{Vincendon2010}  with standard parameters and without subsurface ice; the blue curve represents the stability predicted by our 3D model without including subsurface ice; and the red curve represents the stability predicted by the model including  subsurface ice as described in section \ref{ssec:Ahronsonstudy}. \CO ice deposits observed on pole-facing slopes by CRISM/OMEGA   \cite{Vincendon2010, Vincendon2015} are presented in grey dots. 
    }
    \label{fig:figcompco2distrib}
\end{figure}

We first demonstrate with our new model that the absence of \CO frost on subtropical pole-facing slopes can be explained without the presence of subsurface ice at these latitudes. In  \citeA{Vincendon2010}, the authors used a 1D version of the Mars PCM (without the sub-grid slope parameterization) to study the stability of \CO frost on 30\textdegree~pole-facing slopes. The 1D PCM uses the same physics as the 3D model, but in the 1D, the atmosphere is in radiative equilibrium with the surface studied, and large-scale influence on the local meteorology are not considered. Following \citeA{Vincendon2010}, \CO ice is stable and should be detected by CRISM/OMEGA at a given latitude if the  \CO ice thickness predicted on the slope by the PCM  exceeds hundreds of $\mu$m.  Their predicted stability diagram of \CO frost on steep pole-facing slopes in Eastern Hellas (130-160\textdegree E) is presented in Figure \ref{fig:figcompco2distrib}.   \CO frost observations by  OMEGA (Observatoire pour la Minéralogie, l'Eau, les Glaces et l'Activité) and CRISM (Compact Reconnaissance Imaging Spectrometer for Mars) on  pole-facing slopes reported in \citeA{Vincendon2010} and \citeA{Vincendon2015} are also plotted. \citeA{Vincendon2010} showed that the strong discrepancy between the theoretical stability and the observations, and notably the absence of frost between 22\textdegree S and 34\textdegree S could not be completely explained by varying the ice optical properties, dust opacity, or realistic surface thermal inertia. However, following \citeA{Haberle2008}, they suggested that the most likely explanation for this absence of frost was the presence of buried water ice under these slopes which released heat during the winter and made the \CO ice unstable. This conclusion was  supported by the fact that the observed latitude / solar longitude distribution of \CO ice between 34\textdegree~and 45\textdegree~latitude is narrow, which requires a latitude-dependent heat source, in agreement with the behavior of subsurface water ice with a latitude-dependent depth.

\begin{figure}[ht!]
    \centering
    \includegraphics[width = \textwidth]{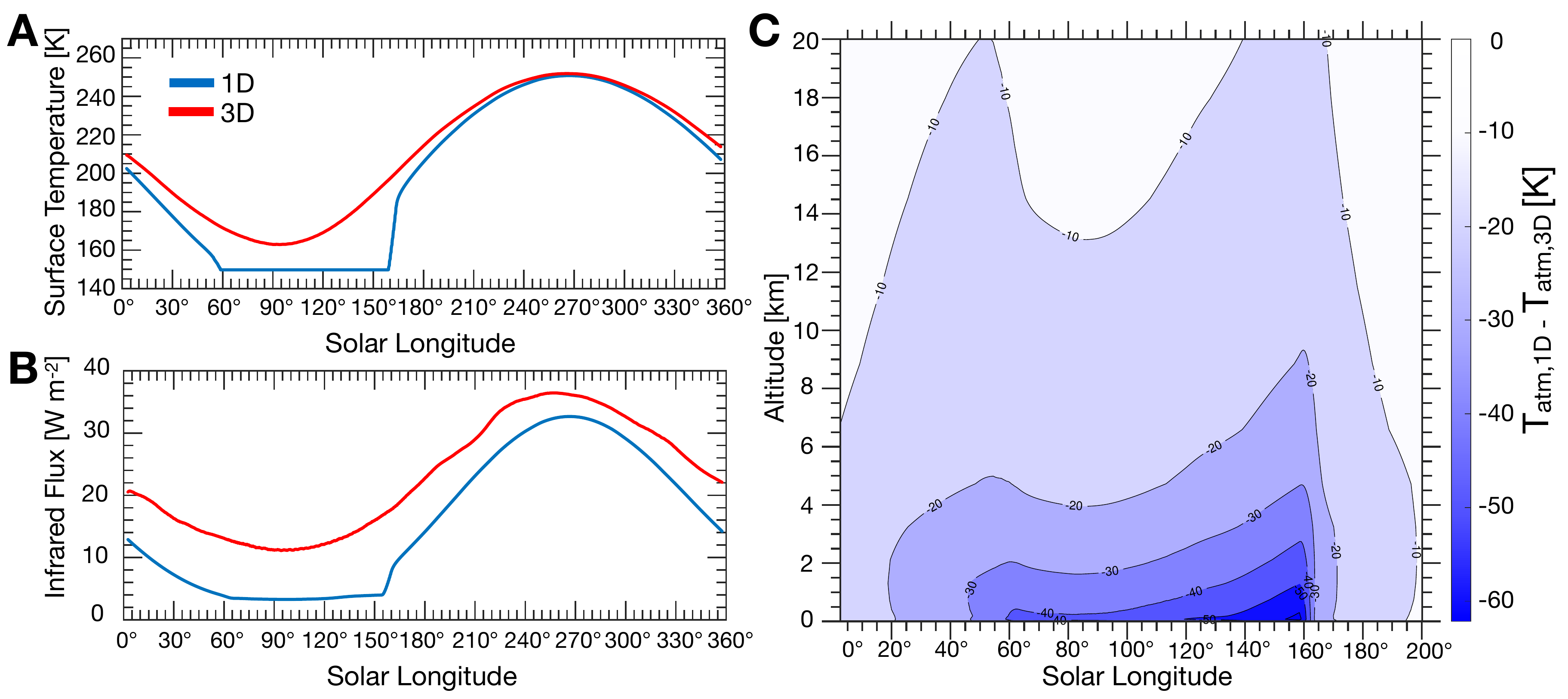}
    \caption{a)  Daily averaged surface temperature of a 30\textdegree~pole-facing slope at 25\textdegree S during the year predicted by the 1D (red curve) and 3D (blue curve) model. b) Same but for the downward infrared flux to the surface. c) Difference of daily averaged atmospheric temperatures between the 1D and 3D models.
    }
    \label{fig:tsurf_fir_tatm}
\end{figure}

We extend their study by computing the theoretical stability of \CO ice on 30\textdegree~pole-facing slopes using our 3D model that simulates the slope microclimates. The same location  and the same criterion for \CO ice stability as in \citeA{Vincendon2010} are used.  First, we experimented without including subsurface ice beneath these slopes. The comparison between the predicted stability from their model and ours is illustrated in Figure \ref{fig:figcompco2distrib}. In our model, \CO frost is never stable for latitudes between $\sim$22\textdegree~and $\sim$33\textdegree~($\pm$3\textdegree~ to account for the grid resolution) contrary to \citeA{Vincendon2010}.  To understand why, we compared the surface temperatures of a 30\textdegree~pole-facing slope at 25\textdegree S computed by the 1D and 3D model (Figure \ref{fig:tsurf_fir_tatm}a). In winter, the surface is 10-15~K warmer in the 3D model compared to the 1D model. We analyzed each of the terms appearing in the surface energy budget (Eq. 1 from Text S1) and found that the major difference between the 1D and the 3D model is in the calculation of the infrared flux (Figure \ref{fig:tsurf_fir_tatm}b). The difference is of the order of 10 to 15~W~m$^{-2}$, which is enough to increase the temperature of a shaded slope by about 10 degrees over most of winter and spring. Indeed, for these surfaces, the solar irradiance around the winter solstice is very low, and the surface temperature becomes very sensitive to the infrared flux.

Two differences  explain this discrepancy:
\begin{enumerate}
    \item 
In the 1D model, the atmosphere is in radiative equilibrium with the studied surface. Above a shaded slope, the atmosphere will be significantly colder than above a flat surface at the same coordinate, even if the slope is very small. In 3D, the atmosphere sees an average of sloped and flat sub-grid surfaces (weighted by  the portion of the grid occupied by these slopes). Steep-sloped terrains actually represent a  small percentage of the overall surface of Mars and are  of limited length  \cite<tens or hundreds of meters,>{Aharonson2006}. Hence, the atmosphere is mostly in equilibrium with the warmer flat sub-grid surface and  the slope microclimates do not significantly impact the state of the emitting  atmosphere. Therefore, the air above the cold sloped surfaces is  warmer in the 3D model compared to the 1D as it is heated by the nearby warm plains. A comparison of the state of the atmosphere between the 1D model with a 30\textdegree~pole-facing slope and our 3D model at the same location is shown in Figure \ref{fig:tsurf_fir_tatm}c. The  difference in the air temperature near the surface (up to $\sim$20-30~km) can be up to 30~or~40~K. This discrepancy is  associated with a difference in the infrared flux of the order of ~10~W~m$^{-2}$, i.e., what is observed in Figure \ref{fig:tsurf_fir_tatm}b. While in our model,  all sub-grid surfaces share the same atmosphere, actually, for a cold slope,  the near-surface atmosphere may tend to cool (or warm) over the first 500~m through radiative exchanges with the surface, and then by convection over the first few kilometers \cite{Read2017}. Hence, the near-surface atmospheric temperature may be colder over a poleward-facing slope than over a flat area.  However, the infrared emission by the atmosphere comes mainly from altitudes between 2 and 10~km \cite<Figure 2 in>{Dufresne2005}. At these altitudes, the atmosphere is not influenced by small slopes \cite<less than 1~km in height difference, as observed on craters where frost is observed, >{Vincendon2010,Vincendon2010water} because it is mixed by winds at altitudes that do not see these small reliefs. For these terrains, the approximation of a shared atmosphere for the calculation of the infrared flux is therefore more relevant.

\item The  1D model does not consider the possible contribution from large-scale meteorology. For example, during winter, the subsidence from the Hadley cell at mid and subtropical latitudes leads to an adiabatic heating of the atmosphere \cite<>[Figure S2]{Read2017}. This warmer atmosphere increases the infrared flux. We quantified this effect by computing the difference of infrared flux reaching a flat surface both in the 1D and 3D model at 25\textdegree S. The computation leads to a difference of $\sim$3-4~W~m$^{-2}$. 

\end{enumerate}

Finally, it should also be noted that a warm atmosphere above a cold slope needs first to be cooled off before condensing on this surface (Text S1). The extra heat brought by the atmosphere when cooled can  nearly reach 2\% of the latent heat, reducing the total mass of \CO condensing, and thus its stability during daytime. When adding these three effects in the 1D model, \CO frost is not expected below 32\textdegree S, as in the 3D model.

Figure \ref{fig:figcompco2distrib} (blue curve) also highlights that our 3D model predicts a delayed condensation and an earlier sublimation than with the 1D model between 45\textdegree~and 35\textdegree~latitude. These effects are not due to the surface albedo/emissivity  between the two models.  We show in Text S2 that the relative time difference between \CO condensation and sublimation for the 1D and 3D model is related to the relative difference in infrared fluxes. This relative difference in fluxes is of a factor of $\sim$1/3 during the southern autumn, and 2/3 in the southern spring (Figure S3a). These ratios are the same as those found for the relative difference in the timing of condensation/sublimation of  \CO  (Figure S3b).

However, we observe that  \CO  condensation occurs later in the observations than predicted by our model without subsurface ice, and earlier for the sublimation. Additionally, the narrow distribution of \CO ice is not well reproduced by our model. This suggests the contribution of a latitude-dependent parameter up to 32\textdegree S \cite<the most equatorial detection by> {Vincendon2015} which is most likely subsurface water ice as demonstrated in \citeA{Vincendon2010}. Considering the resolution of our model, we conclude that above 30\textdegree S, subsurface ice is required to explain the narrow distribution of \CO ice, but is not necessary to explain the absence of \CO ice equatorward.

\subsection{Theoretical stability of subsurface ice beneath pole-facing slopes \label{ssec:Ahronsonstudy}}

\begin{figure}[ht!]
    \centering
    \includegraphics[width = 1.2\textwidth]{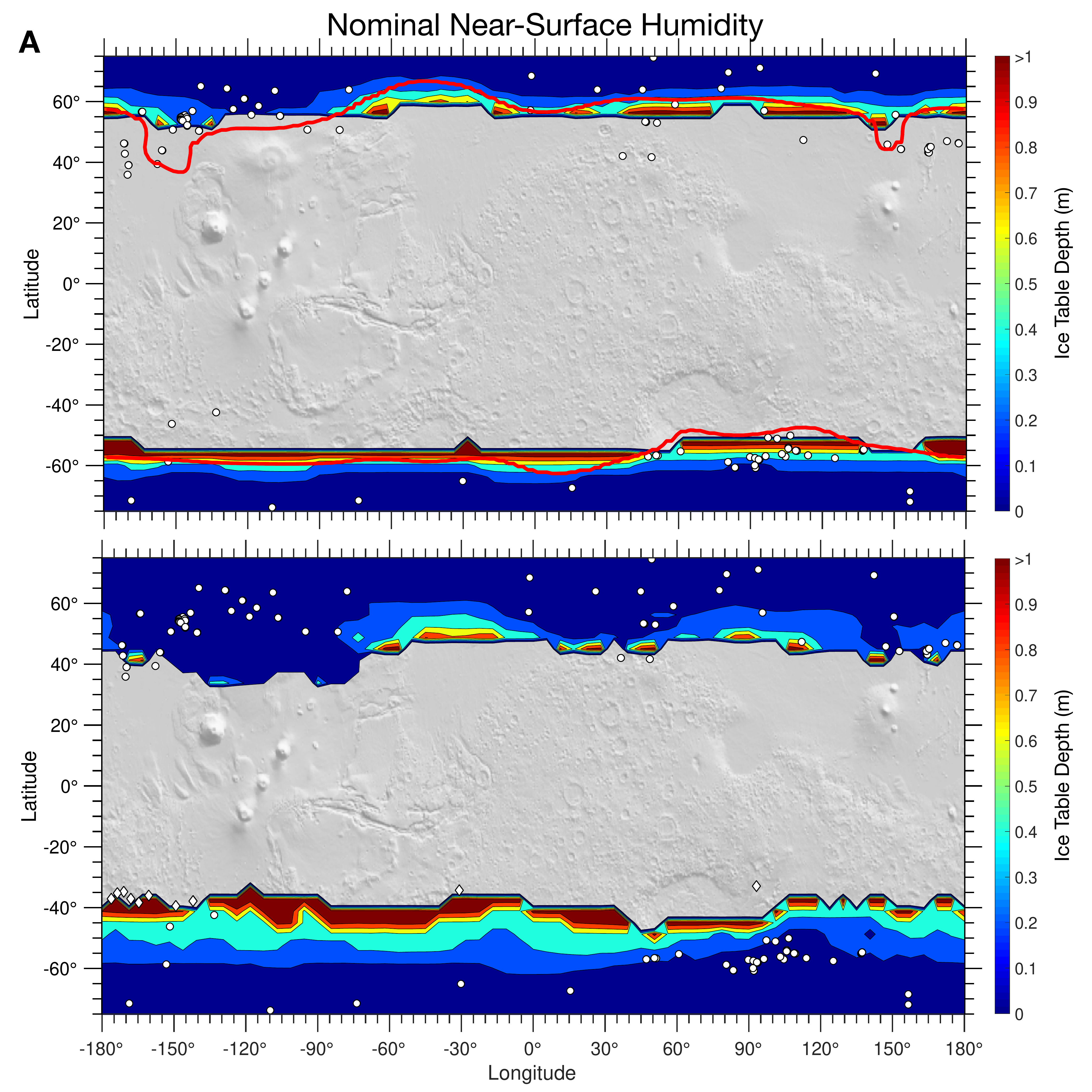}
    \caption{Theoretical stability of subsurface water-ice with respect to diffusion for a) flat surfaces b) 30\textdegree~pole-facing slopes, using a nominal near-surface humidity. The red curve is the observed 10\% Water-Equivalent Hydrogen contour, which is a good proxy for the presence of water ice in the shallow subsurface \cite{PATHARE2018}. White dots indicate  exposed water ice along cliff scarps and impact craters as reported in \citeA{Dundas2021ice, Dundas2022} White diamonds indicate exposed water ice along gullies \cite{Khullericegullies}. 
    }
    \label{fig:icetable}
\end{figure}

We now investigate the possible stability of subsurface ice on subtropical slopes following the approach described in Section \ref{ssec:SSI}. First, we test our method by calculating the stability of water ice under flat terrain (Figure \ref{fig:icetable}). Subsurface ice is stable according to our model poleward of 55\textdegree, with brief excursions to 50\textdegree~in regions of high albedo and low thermal inertia. Overall, the simulated spatial distribution of ice is consistent with MONS measurements \cite{DIEZ2008, PATHARE2018} and surface ice exposures \cite{Dundas2021ice}. However,  some differences exist (e.g., MONS predicts ice down to latitudes 40\textdegree N, 45\textdegree S at some longitudes, and some recent impact craters reveal subsurface ice \cite{Dundas2021ice} where it is not predicted by our model. These differences are discussed in section \ref{sec:Discussion}. In addition, the depths at which ice is stable are  broadly consistent with those published in the literature \cite{Mellon2004, Schorghofer2005, Chamberlain2007, DIEZ2008, PATHARE2018, Piqueux2019} (Figure S4). At the Phoenix landing site, ice is predicted to be stable at a depth of 9~cm, in good agreement with direct in situ observations which reveal a depth between 5 and 18~cm  \cite{Smith2009}. 

We then model the possible stability of ice underneath pole-facing terrain with a slope angle of 30\textdegree. The results are presented in Figure \ref{fig:icetable}b. The distribution of stable ground-ice extends 
equatorward, with limits up to $\pm 35$\textdegree. Excursions to lower latitudes are located in areas of low thermal inertia and high albedo (e.g., East Hellas, East Tharsis) and do not extend below 30\textdegree~of latitudes. Our model thus differs from those of \citeA{Aharonson2006} and of \citeA{Mellon2022} which predicted stable ice down to 25\textdegree~latitude.

Two main differences between  their models and ours can explain our more limited latitudinal  extent  for subsurface ice.  First, in \citeA{Aharonson2006}'s model, the infrared flux is computed as 4\% of the solar flux at noon. \citeA{Haberle1991} showed that approximating the infrared flux as 2\% of the solar flux at noon resulted in an underestimation of the infrared flux. Our calculations show that even if one replaces the 2\% by 4\% as in \citeA{Aharonson2006}, the infrared flux is still underestimated by about 2 to 6~W~m$^{-2}$ at subtropical latitudes with low dust opacity, and up to tens of W~m$^{-2}$ for dusty periods.  \citeA{Mellon2022}'s model is a priori less sensitive to this last effect since the infrared flux is computed with the atmospheric model of \citeA{Pollack1990}. Differences may occur since our model has improved the radiative treatment of dust and clouds \cite{Madeleine2011, Madeleine2012}, and directly takes  into account the dust observations by \citeA{MONTABONE2015}. Finally, in  \citeA{Mellon2022, Aharonson2006}'s models, $p_{vap,surf}$ in Eq. \ref{eq:stabilityssi} is computed with surface humidity from column-integrated measurements by TES obtained during daytime  \cite{Smith2002} if the surface is not at saturation. In our model, $p_{vap,surf}$  is computed by the Mars PCM that solves the complete diurnal and seasonal water cycle  and the vertical diffusion. Hence, as we are  considering the complete diurnal cycle of water vapor (v.s. daytime measurements for \citeA{Aharonson2006,Mellon2022}), our surface humidity might be lower compared to the other models. Furthermore, the calculation of surface humidity from column-integrated measurements is complex because the vertical structure of the water vapor at the near-surface is not very constrained \cite{Tamppari2020}. To date, their models assume a well-mixed, hydrostatic, and isothermal atmosphere \cite{Schorghofer2005}.   The difference between the predicted near-surface humidity from PCM and that obtained from  TES daytime measurement interpolation can reach up to 0.05~Pa, i.e, the humidity in \citeA{Aharonson2006,Mellon2022} can be  20\% higher than of the humidity retrieved with the PCM. Such a difference is significant for near-equatorial regions, where the stability of subsurface ice depends essentially on the near-surface water content \cite{SONG2023}. Finally, we note that in \citeA{Mellon2022}'s model, the humidity used is 2.6 times that observed by TES, which tends to increase the stability of the near-equatorial ice in their model. This last point is discussed in section \ref{ssec:unstableice}.

\section{Discussions \label{sec:Discussion}}

\subsection{The sparse presence of modeled subsurface water ice below $\pm$30\textdegree N \label{ssec:sparsesurfaceconditions}}
\citeA{Vincendon2010} reported that water ice should be present within one meter of the surface on all 20–30\textdegree pole-facing slopes down to about 25\textdegree S. They predicted that subsurface ice may be stable  even at equatorward latitudes ($\sim 20$\textdegree S) with favorable slope conditions (very steep dusty slope). We investigate here the sensitivity of our model to the parameters used in this study. For instance, for very steep slopes (40\textdegree), our model suggests the presence of \CO ice down to 27\textdegree S. Even if these slopes are sparse \cite{Aharonson2006}, no \CO ice detections have been made on such slopes at latitudes lower than 32\textdegree S \cite{Vincendon2010, Vincendon2015}, suggesting the presence of water ice in these rare locations. \citeA{Vincendon2010} also showed the sensitivity of \CO ice formation/sublimation to surface properties.  No clear constraints exist for the slope surface properties as some slopes exhibit low thermal inertia \cite{Tebolt2020}, favoring the condensation of \CO; and some slopes reveal high-thermal inertia bedrock exposures \cite{Edwards2009}  which inhibit the formation of \CO frost. Yet, in the most favorable case (high albedo, low thermal inertia), our model suggests the presence of \CO ice down to 25\textdegree S where no ice is observed \cite{Vincendon2010, Vincendon2015}. Hence, this suggests that water ice should be present beneath steep pole-facing slopes down to $\pm$30\textdegree~of latitudes on average, and could be present down to 25\textdegree S for sparse locations with favorable conditions (steep slopes $\geq$ 40\textdegree, high albedo, low thermal inertia).

Our model was validated by comparison with surface temperatures measured on sloped terrain and seasonal variations in water frost formation \cite{Lange2023Model}. It turned out that our model could overestimate certain temperatures by 2~K on average, and up to 5~K on certain poleward-facing slopes, depending on local terrain properties (thermal inertia, slope angle, azimuth). If such a positive bias were confirmed, it could mean that ice stability would extend to 28$^\circ$ latitude for some 30\textdegree~slopes (and 23\textdegree~for some 40\textdegree~slopes).

\subsection{Possible presence of unstable water ice \label{ssec:unstableice}}

The subsurface ice model used previously only allows us to determine the depth at which diffusion-formed subsurface water ice can be stable and in equilibrium with the atmosphere. According to our model, pore-filling water ice is stable down to latitudes of about 55\textdegree~and locally 52\textdegree.  However, MONS measurements \cite{DIEZ2008, PATHARE2018} have shown that water ice is expected underneath horizontal surfaces down to latitudes below 45\textdegree~at Arcadia and Utopia Planitia, where our model does not predict stable ice (Figure \ref{fig:icetable}). Seasonal variations in surface temperatures monitored by the Mars Climate Sounder also indicate traces of near-surface ice down to depths of less than 1~m at latitudes of 45\textdegree~\cite{Piqueux2019}. Finally, ice excavations at impact craters show near-surface ice down to latitudes of 35\textdegree N \cite{Byrne2009,Dundas2014, Dundas2021ice, Dundas2022}. These exposed ice chunks may be more like pure ice than pore-filling due to the low regolith content in the ice \cite{Dundas2014, Dundas2021ice, Dundas2022}. In each case, our model, as well as those of \citeA{Mellon2004, Schorghofer2005, Chamberlain2007}  do not predict stable ice at these locations with current humidity (Figure \ref{fig:icetable}). 

To solve this paradox, \citeA{Mellon2004,Chamberlain2007,Byrne2009}  doubled the humidity in their model to fit the MONS observations. Such a calculation assumes that the observed stable ice distribution is representative of that of the last several thousand years, where the authors assume the global average column abundance was at least twice as high as the 10 pr-$\mu$m observed today \cite{Smith2002}. \citeA{Schorghofer2012, Bramson2017}  proposed instead that these ices are traces of former ice formed as a result of past obliquity variations \cite<e.g., >{Levrard2004,Madeleine2009,Madeleine2014}, which are subsequently protected by the formation of a lag deposit. Thus, the subsurface ice observed today would not be in equilibrium with the surface and could act as a source of water vapor today \cite{Schorghofer2012}. Following \citeA{Mellon2004,Chamberlain2007,Byrne2009} approaches, we found that we needed to triple the near-surface humidity to find similar subsurface ice distribution (Figure S5a). Note that in this extreme case, the average latitudinal extent of subsurface ice exceeds the average excess-ice limit  observed by MONS.

The same  question arises for the ice underneath pole-facing slopes where old unstable ice could persist. Possible direct observations of subsurface ice have been reported at 32.9\textdegree S \cite{Khullericegullies} and geomorphic traces linked to the presence of ice in the subsurface have been detected down to latitudes of 30\textdegree~\cite{Viola2018} whereas our model predicts stable subsurface ice down to latitudes of 35\textdegree~at depths of the order of a meter. We test the sensitivity of this result to the surface humidity conditions by tripling the near-surface humidity for the flat terrains. The distribution obtained is presented in Figure S5b. In this extreme scenario, subsurface ice is predicted to be stable down to $\pm$30\textdegree N, with brief excursions down to $\pm$25\textdegree N in favorable areas (high albedo, low thermal inertia). Hence, this experiment reinforces the conclusions drawn in section \ref{ssec:sparsesurfaceconditions}, i.e., that water ice could be present beneath steep pole-facing slopes down to $\pm$30\textdegree~of latitudes on average, and $\pm$25\textdegree~locally.

Our slope microclimate model without subsurface ice (Figure \ref{fig:figcompco2distrib}, blue curve) starts condensing \CO  too early (by about 10\textdegree~of~\Ls) compared to frost observations. By introducing subsurface ice at a depth given in Figure \ref{fig:icetable}, we find that the new distribution for \CO frost stability differs only very slightly from that without subsurface ice (Figure \ref{fig:figcompco2distrib}, red curve). This result is expected because the ice is at depths greater than the thermal skin thickness associated with the seasonal cycle and thus does not have a strong impact on the surface energy budget. Hence, the narrow distribution of \CO ice observations for latitudes higher than  30\textdegree S requires subsurface ice with shallower latitude-dependent depths than those predicted by our model, as reported in \citeA{Vincendon2010}. Even with the depths obtained from the triple humidity case, we can not correctly fit the narrow distribution, suggesting the presence of shallower (and thus unstable) ice.  Future work will investigate the formation of glaciers during past epochs, the formation of lag deposits during their sublimation period, and their current preservation. Local measurements of seasonal variations of  surface temperatures on sloped terrains could also help to constrain the presence of subsurface ice at low latitudes \cite<e.g.,>{Bandfield2007, Piqueux2019}.

\section{Conclusions \label{sec:conlusion}}
During this study, we have extended the work of \citeA{Vincendon2010}, \citeA{Aharonson2006} and \citeA{Mellon2022}  who proposed the presence of subsurface water ice below 30\textdegree~of latitude. On one hand, in \citeA{Vincendon2010}, the absence of \CO frost on subtropical slopes was linked to the presence of high thermal inertia subsurface water ice that released heat during winter, preventing  \CO condensation. Here, our model of slope microclimate shows that  \CO ice is unstable on most  slopes  without subsurface water ice. Indeed, the plains surrounding the slope heat the atmosphere, increasing the infrared flux reaching the slope, warming the surface, and preventing it from reaching the  \CO condensation temperature. On the other hand, the subsurface ice stability model from \citeA{Aharonson2006} predicted stable ice down to 30\textdegree~of latitude underneath pole-facing slopes, and down to 25\textdegree~of latitude in dusty areas for the steepest slopes. Our subsurface ice stability model, coupled with the slope microclimate model, shows that slopes at these latitudes are too warm for stable subsurface ice with the current humidity and that this ground ice is only stable poleward of 30\textdegree.  Our study reappraises this latitudinal extent of water ice proposed in these studies: subsurface water ice should be present beneath steep ($\geq$30\textdegree) pole-facing slopes down to 30\textdegree~of latitudes on average, with sparse excursions down to 25\textdegree~for favorable locations (steep slopes, high albedo, low thermal inertia).  However,  subsurface stability models cannot conclude definitively about the presence of ice, since it does not model unstable ice remaining from past ice ages. Several markers suggest the possible presence of vestige unstable subsurface ice at low latitudes \cite{Dundas2014, Dundas2021ice, Viola2018}. Similarly,  for latitudes above 32\textdegree S, our model does not exactly reproduce the narrow distribution of CO$_2$, suggesting the presence of shallower unstable ice. Modeling the accumulation, burial, and preservation of this ice during glacial periods, as well as more observational constraints on these near-equatorial slopes, will allow us to accurately characterize the presence or absence of subsurface ice.  Our study suggests that water ice resources would be thus sparse at latitudes lower than 30\textdegree. Therefore, the accessibility of other water reservoirs like hydrated minerals should be more characterized at these latitudes as part of the strategy to rely on In Situ Resources for future crewed Martian missions.

\section{Open Research}
\CO frost detections by OMEGA and CRISM are from \citeA{Vincendon2010,Vincendon2010water,Vincendon2015}.  Data files for figures used in this analysis are available in a public repository, see \citeA{LANGEDATASET}. The Mars PCM  used in this work can be downloaded with documentation from the SVN repository at \url{https://svn.lmd.jussieu.fr/Planeto/trunk/LMDZ.MARS/.} More information and documentation are available at \url{https://www-planets.lmd.jussieu.fr}.


\acknowledgments
This project has received funding from the European Research Council (ERC) under the European Union’s Horizon 2020 research and innovation program (grant agreement No 835275, project "Mars Through Time").  Mars PCM simulations were done thanks to the High-Performance Computing (HPC) resources of Centre Informatique National de l’Enseignement Supérieur (CINES) under the allocation n\textdegree~A0100110391 made by Grand Equipement National de Calcul Intensif (GENCI). The authors thank Editor K. Lewis, M. Kreslavsky, and an anonymous reviewer for their comments. The authors thank  N. Schörghofer, C. Dundas, A. Khuller for  insightful discussions.

\nocite{Christensen2004,deVrese2016,Pitman2003,Wang2016}
\bibliography{agusamble}

\end{document}


%
%


\title{Supporting Information for "A Reappraisal of  Near-Tropical Ice Stability on Mars"}
%
%

%
%



\authors{L.Lange\affil{1}, F.Forget\affil{1}, M.Vincendon\affil{2}, A.Spiga\affil{1}, E.Vos\affil{1,3}, O.Aharonson\affil{3},
E.Millour\affil{1}, R.Vandemeulebrouck\affil{1},  A.Bierjon\affil{1}
}


\affiliation{1}{Laboratoire de Météorologie Dynamique,Institut Pierre-Simon Laplace (LMD/IPSL), Sorbonne Université, Centre National de la Recherche Scientifique (CNRS), École Polytechnique, École Normale Supérieure (ENS), Paris, France}
\affiliation{2}{Institut d’Astrophysique Spatiale, Université Paris-Saclay, CNRS, Orsay,France}
\affiliation{3}{Department of Earth and Planetary Sciences, Weizmann Institute of Science, Rehovot, Israel}

%
%

%
\begin{article}

%
%

\noindent\textbf{Contents of this file}
\begin{enumerate}
\item Text S1 to S2
\item Figures S1 to S5
\end{enumerate}

\noindent\textbf{Introduction}
This supporting information document contains two sections. The first section summarizes the slope modeling approach described in \cite{Lange2023Model}, and the second section demonstrates the relationship between the variation of the infrared flux and the  variation of the timing of condensation/sublimation of CO$_2$ frost.

Five figures are included to support the text and supporting text mentioned here above: 
\begin{enumerate}
    \item A cartoon describing the sub-grid scale parameterization to model slope microclimate in the Mars Planetary Climate Model;
    \item A map of zonal mean atmospheric temperature during Southern solstice to highlight the subsidence of the atmosphere during winter;
    \item A figure illustrating the relative difference in the infrared flux, and its correlation with the timing of condensation/sublimation of CO$_2$ frost
    \item A comparison between the depth of the subsurface ice obtained from our model v.s. other models/measurements. 
    \item The distribution of subsurface ice obtained when tripling the near-surface humidity.
\end{enumerate}

\begin{figure}[!ht]
    \centering
    \includegraphics[width = \textwidth]{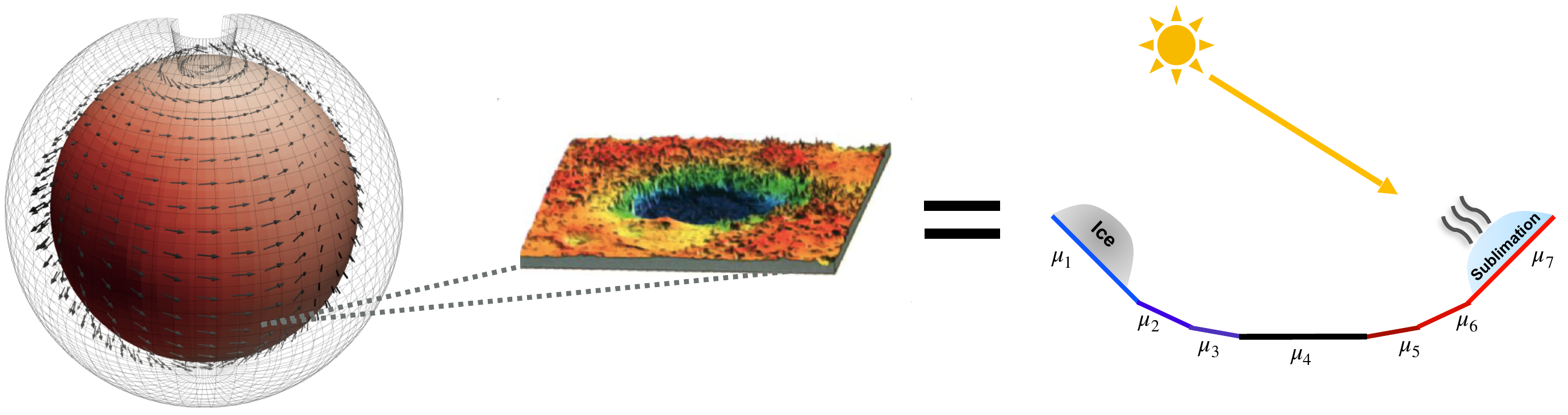}
    \caption{Schema of the sub-grid scale slope parameterization. Each mesh of the Mars PCM is decomposed into sub-grid slopes (defined by characteristic slopes $\mu_i$) or flat terrain. These sub-grid terrains have their microclimate and the interactions between the atmosphere and surface are made through averaged values over the mesh.  }
    \label{fig:cartoon_sub-gridslopes}
\end{figure}

\noindent \textbf{Text S1: Mars Planetary Climate Model with sub-grid slope parameterization \label{Supp:model}}
We present here the development of a sub-grid slope parameterization in the Mars Planetary Climate Model to model sub-grid slope microclimates. A complete description and validation of the model can be found in \citeA{Lange2023Model}.  Inspired by \textit{Land Surface Parameterization}  strategies used in Earth Climate Model to consider sub-grid surface heterogeneities \cite<e.g.,>{Pitman2003,deVrese2016},  we have developed a  sub-grid scale parameterization that accounts for topographic heterogeneity in the Mars Planetary Climate Model. The principle is illustrated in  Figure. \ref{fig:cartoon_sub-gridslopes}. For each GCM mesh, we decompose the cell as a distribution of sloped terrains (defined by characteristic slopes) and a flat terrain. The sloped terrains are either North or South-facing: we have shown in \citeA{Lange2023Model} that any slopes of angle $\theta$ and azimuth $\psi$ can be thermally represented by a North or South facing slope of slope angle  $\theta \cos(\mu)$. Seven sub-grid surfaces are considered. 

Then, on each sub-grid surface, we compute the surface temperature by solving: 
\begin{equation}
   \rho  c_p \frac{\partial T_{\rm{surf,slope}}}{\partial t} = F_{\rm{rad,slope}}(t) + F_{\rm{ground,slope}}(t) +  F_{\rm{atm,slope}}(t) + \sum_{i}L_i\frac{\partial m_i}{\partial t}   - \epsilon \sigma T_{\rm{surf,slope}}^4(t)
   \label{eq:tsurfsolve}
\end{equation}

\noindent where $\rho$ is the surface density of the ground (kg~m$^{-2}$), $c_p$ is the specific heat capacity (J~K$^{-1}$),  $T_{\rm{surf,slope}}$ the sub-grid slope temperature (K), $F_{\rm{rad,slope}}$ the radiative flux at visible and infrared wavelengths (W~m$^{-2}$), $F_{\rm{ground,slope}}$ the geothermal flux (W~m$^{-2}$), $F_{\rm{atm,slope}}$ the sensible heat flux, $\sum_{i}L_i\frac{\partial m_i}{\partial t} $ the latent heat flux due to the condensation/sublimation of a volatile with a latent heat $L_i$ (J~kg$^{-1}$) and $ \epsilon \sigma T_{\rm{surf,slope}}^4$ the radiative cooling of the surface where $\epsilon$~(1) is the surface emissivity, $\sigma$~(W~m$^{-2}$~K$^{-4}$) the Stefan-Boltzmann constant.

In the visible wavelengths, $F_{\rm{rad,slope}}$ is computed following \citeA{Spiga2008}. In the infrared, assuming that the thermal emission is isotropic, and based on geometric considerations, $ F_{\rm{IR, slope}}$ is computed with:
\begin{equation}
     F_{\rm{IR,slope}} = \sigma_s  F_{\rm{rad~IR,flat}} + (1-\sigma_s)\epsilon_{\rm{flat}}\sigma T_{\rm{surf,flat}}^4
     \label{eq:eqFir}
\end{equation}

\noindent where $\sigma_s    = \frac{1+\cos(\theta)}{2}$ (1) is the sky-view factor, $F_{\rm{IR,flat}}$ (W~m$^{-2})$ the atmospheric incident thermal IR flux on a horizontal surface computed by the PCM,  $\epsilon_{\rm{flat}}$ (1) is the emissivity sub-grid flat terrain,  $T_{\rm{surf,flat}}$ (K) its temperature.

The sensible heat flux is given by:
\begin{equation}
    F_{\rm{atm}} = \rho C_d U(T_{\rm{atm, 1}} - T_{\rm{surf}})
\end{equation}
\noindent where  $U$ (m~s$^{-1}$) is the wind velocity,  $T_{\rm{atm, 1}} $ (K) the temperature of the atmosphere in the 1$^{st}$ vertical layer of the GCM about 4~m above the surface and $C_d$ (1) is a drag coefficient.

Here, we assume that the sub-grid surfaces share the same atmosphere and turbulent exchanges are only computed with the flat sub-grid surface. This is justified because 1) in each of the PCM cells, the sub-grid flat terrain is predominant and slopes are sporadic \cite{Lange2023Model}; 2)  The slopes modeled here have a low extent  (a few kilometers) compared to the typical values of winds ($\sim$ meters per second). Hence, winds homogenize the air above the sub-grid surfaces. A complete description of turbulent exchanges through each sub-grid surface would require resolving the slope winds, which are not handled by our model yet.

On each sub-grid terrain, we let the microclimates evolve so that model slope-specific phenomena  (e.g., condensation of volatiles, formation of glaciers, migration of subsurface ice, etc.) can be simulated. The portion of the atmosphere above the ground within the cell sees an average of these surface microclimates.

For each sub-grid surface, water ice deposition and sublimation are computed following \citeA{Navarro2014}.  CO$_2$ deposition and sublimation computations and the latent heat associated are made following \citeA{Forget1998snow}. The amount of CO$_2$ condensing $\delta m$ (kg~m$^{-2})$ is:

\begin{equation}
    \delta m = \frac{c_p}{L_{CO_2}+ c_{\rm{air}}(T_{\rm{atm, 1}} - T_{\rm{cond}})}\left(T_{\rm{cond}} - T_0 \right)
\end{equation}

\noindent where $L_{CO_2}$ (J~kg$^{-1}$) is the latent heat of CO$_2$, $c_{\rm{air}}$ (J~kg$^{-1}$~K$^{-1}$) is the air specific heat at constant pressure, $T_{\rm{cond}}$~(K) the condensation temperature  at surface pressure . This expression is almost similar to the one of  \citeA{Forget1998snow} except we add the term  $c_{\rm{air}}(T_{\rm{atm, 1}} - T_{\rm{cond}})$  which corresponds to the extra heat brought by the atmosphere when cooled to the condensation temperature $T_{\rm{cond}} $ just above the surface.

Our model has been  validated through comparison with surface temperatures measurements by the Thermal Emission Spectrometer  \cite<TES,>{Christensen2001} and  Thermal Emission Imaging System \cite<THEMIS,>{Christensen2004}, and with comparisons of the temporal and spatial distribution of surface frost  (both water and CO$_2$) predicted by our model v.s. observed by CRISM and OMEGA \cite{Vincendon2010,Vincendon2010water}. 

\noindent\textbf{Text S2.}

The rate of sublimation/condensation of a CO$_2$ frost with a thickness $\delta$ (m) is given by:

\begin{equation}
    \rho L \frac{\partial \delta }{\partial t} = F_{\rm{IR}} - \epsilon \sigma T_{CO_2}^4 + (1-A) F_{\rm{vis}} + F_{\rm{cond}}+ F_{\rm{turb}}
    \label{eq:bilanCO2}
\end{equation}

\noindent where $\rho$ is the ice density (kg~m$^{-3}$), $L$ the latent heat of sublimation (J~kg$^{-1}$),  $F_{\rm{IR}}$ the infrared flux, $T_{CO_2}$ (K) the temperature of condensation for CO$_2$ , $F_{\rm{vis}}$  the visible flux, $F_{\rm{cond}}$  the geothermal flux, $F_{\rm{turb}}$ the turbulent flux (all fluxes in W~m$^{-2}$); $A$ the albedo (1) and $\epsilon$ the emissivity (1).  As we are interested in the sensitivity of the condensation/sublimation rate with regard to the infrared flux computed by the 1D and 3D models, we neglect the influence of the last three terms. Comparisons between the 1D and 3D models show that $F_{\rm{vis}}$ was similar in both cases, and $F_{\rm{turb}}$  can be neglected compared to $F_{\rm{IR}}$ as well as $F_{\rm{cond}}$ in the absence of subsurface ice.  $F_{\rm{vis}}$ can indeed be neglected for areas in the polar night or pole-facing slopes that do not receive a significant amount of solar energy during late autumn/winter. Hence,  Eq. \ref{eq:bilanCO2} can be approximated as:

\begin{equation}
    \rho L \frac{\partial \delta }{\partial t} = F_{\rm{IR}} - \epsilon \sigma T_{CO_2}^4 
    \label{eq:bilanCO2_simplify}
\end{equation} 

The net budget at infrared wavelength is noted:

\begin{equation}
\tilde{F_{IR}}  = F_{IR} - \epsilon \sigma T_{CO_2}^4 \label{eq:netinfraredbudget}
\end{equation} 

From Eq. \ref{eq:bilanCO2_simplify},  \ref{eq:netinfraredbudget}, it can be approximated that the time $\tau$ to sublimate a frost with thickness $\delta$ is:
\begin{equation}
\tau = \frac{\rho L \delta}{\tilde{F_{IR}} }
\label{eq:tausublimation}
\end{equation} 

We now derive the relative difference $\Delta \tau$ for $\tau$ when computing the infrared flux with the 1D and the 3D model. By differentiating Eq. (3), and assuming that $T_{CO_2}$ is the same in both cases, we obtain:

\begin{equation}
\frac{\Delta \tau }{\tau^2} \sim \frac{\Delta \tilde{F}_{IR}}{\rho \delta L} \label{eq:relativetau}
\end{equation}

And using Eq. \ref{eq:tausublimation}:

\begin{equation}
\frac{\Delta \tau }{\tau } \sim \frac{\Delta \tilde{F}_{IR}}{\tilde{F}_{IR}}
\label{eq:finalrelativetau}
\end{equation}

\begin{figure}[!ht]
 \noindent\includegraphics[width=\textwidth]{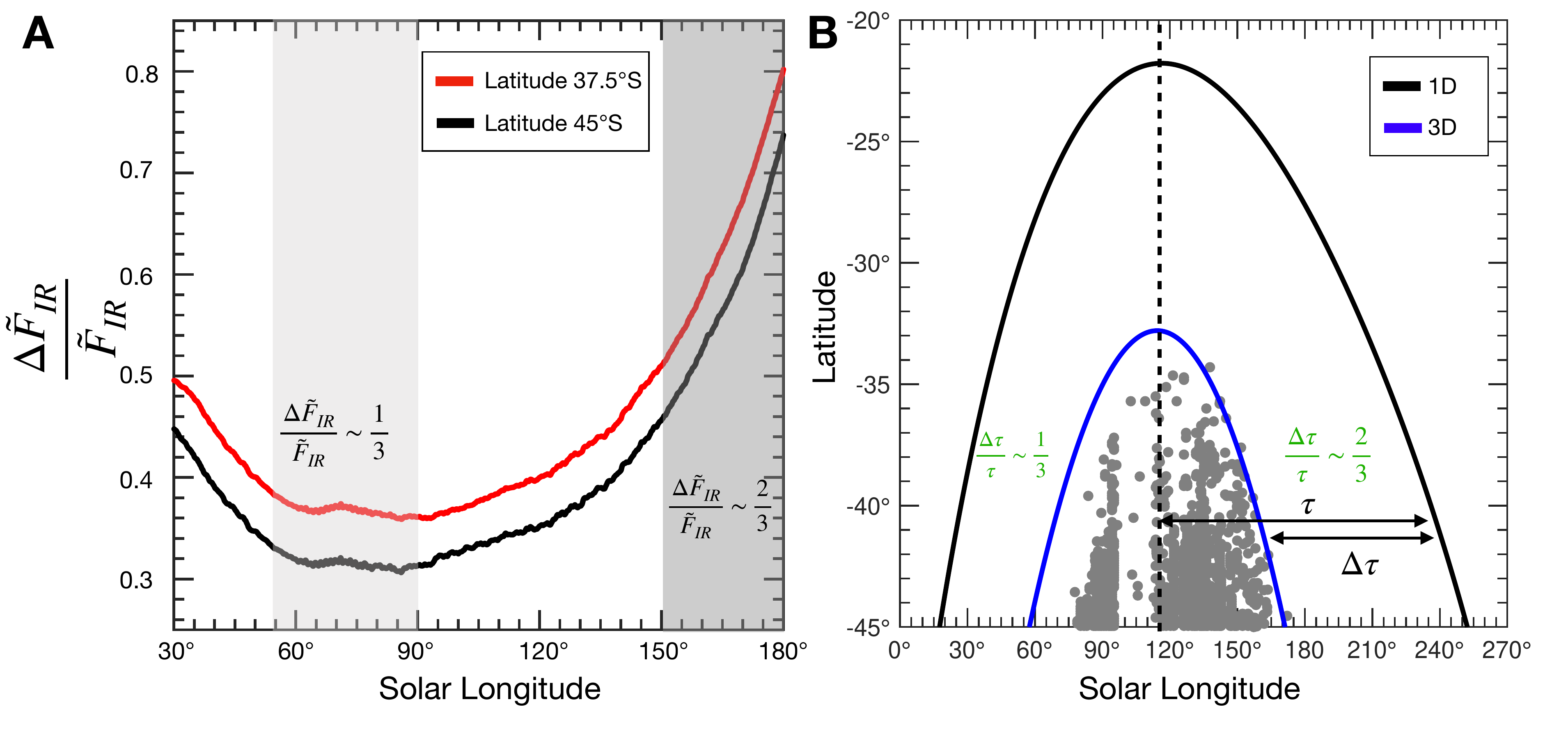}
\caption{a) Evolution of $ \frac{\Delta \tilde{F}_{IR}}{\tilde{F}_{IR}} $ with the solar longitude for the latitudes of 37.5\textdegree S (red curve) and  45\textdegree S b) Predicted stability of CO$_2$ on 30\textdegree~pole-facing slopes for the 1D (dark curve) and the 3D model (blue curve). $\tau$ indicates the duration of condensation/sublimation for the 1D model, $\delta \tau$ is the difference in the duration of condensation/sublimation between the 1D and 3D model.  }
\label{fig:relFir_relTau}
\end{figure}

We have computed the relative difference of $\tilde{F}_{IR}$ between the 1D and 3D over the winter. Results are presented in Figure. \ref{fig:relFir_relTau}a. According to this computation, the relative difference of $\tilde{F}_{IR}$ is about  $\sim$30\% during southern autumn, and $\sim$60\% during spring. Hence, according to Eq. \ref{eq:finalrelativetau},  the condensation should be delayed by 30\%, and sublimation should be advanced by 60\%. This is consistent with what is observed when comparing the 1D and 3D models that predicted the start of condensation and end of sublimation (Figure. \ref{fig:relFir_relTau}b).

 \bibliography{agusamble}

%
%
%
%
%

%
%
\end{article}
\clearpage

\begin{figure}[!ht]
    \centering
    \includegraphics[width = \textwidth]{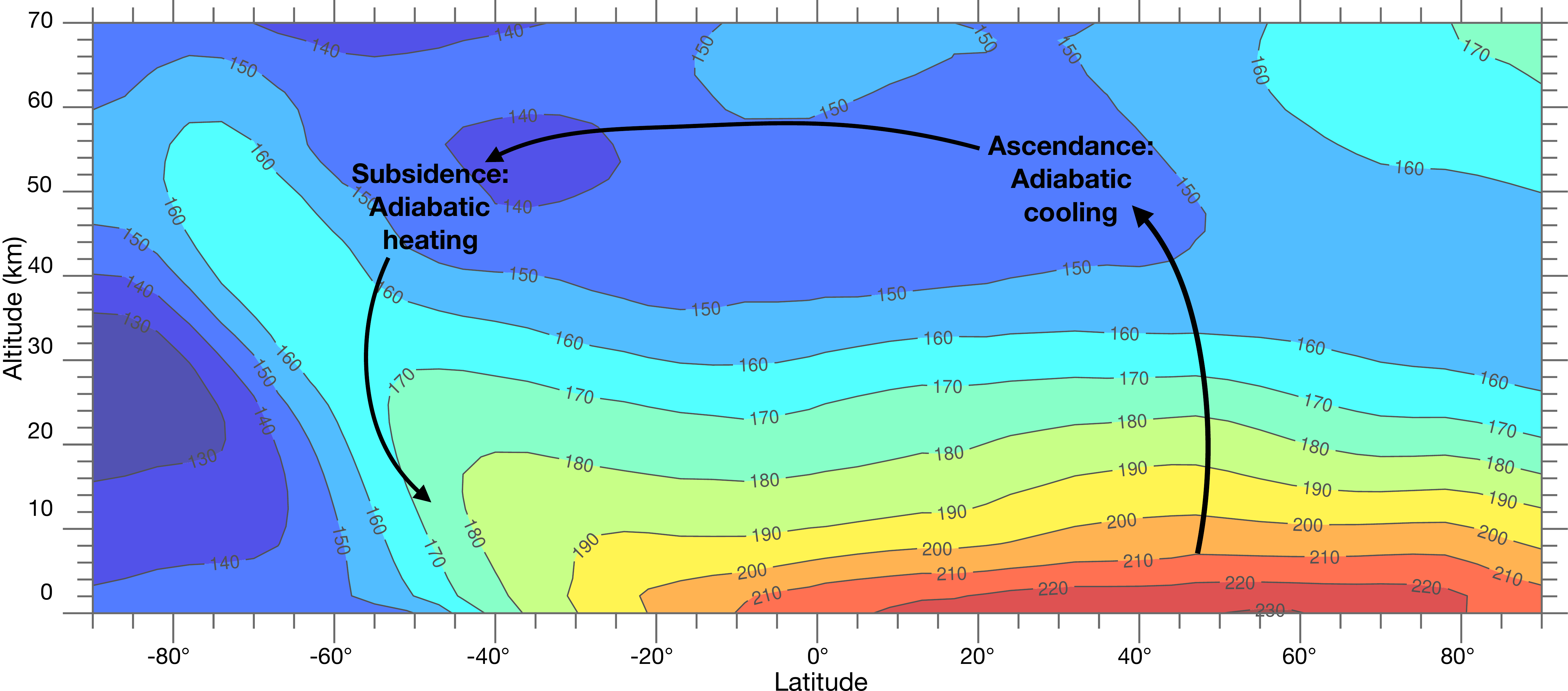}
    \caption{Zonal mean of atmospheric temperature during southern solstice (L$_s$~=~90\textdegree). Atmospheric temperatures are obtained from the Mars PCM. }
    \label{fig:subsidence}
\end{figure}

 \begin{figure}
 \noindent\includegraphics[width=\textwidth]{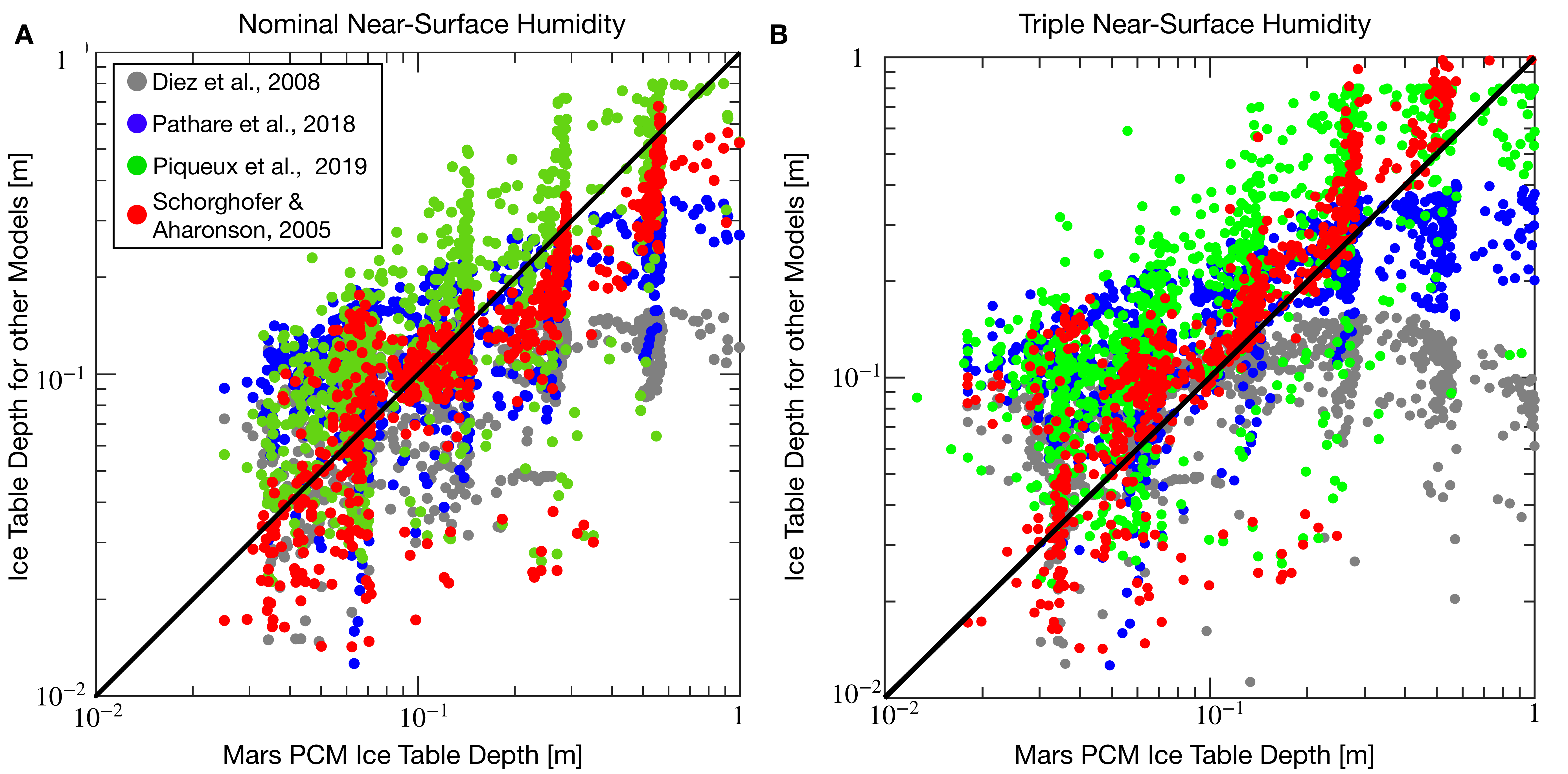}
\caption{Comparison of the depth of the ice table beneath a flat surface predicted by the PCM with several models and data for a) the nominal near-surface humidity computed by the PCM b) a triple near-surface humidity. The grey dots correspond to the ice table depth obtained from MONS measurements by \citeA{DIEZ2008}; blue dots are depth computed from MONS measurements by \citeA{PATHARE2018}; green dots are depth obtained from thermal measurements by \citeA{Piqueux2019}; red dots are the predicted depth by \citeA{Schorghofer2005} vapor diffusion model. For \citeA{DIEZ2008,PATHARE2018,Schorghofer2005}, depths are given in g~cm$^{-2}$. We assume a regolith density of 1.5~g~cm$^{-3}$ to convert it to a depth in cm.  }
\label{fig:figcompicetabledepth}
\end{figure}

 \begin{figure}
 \noindent\includegraphics[width=\textwidth]{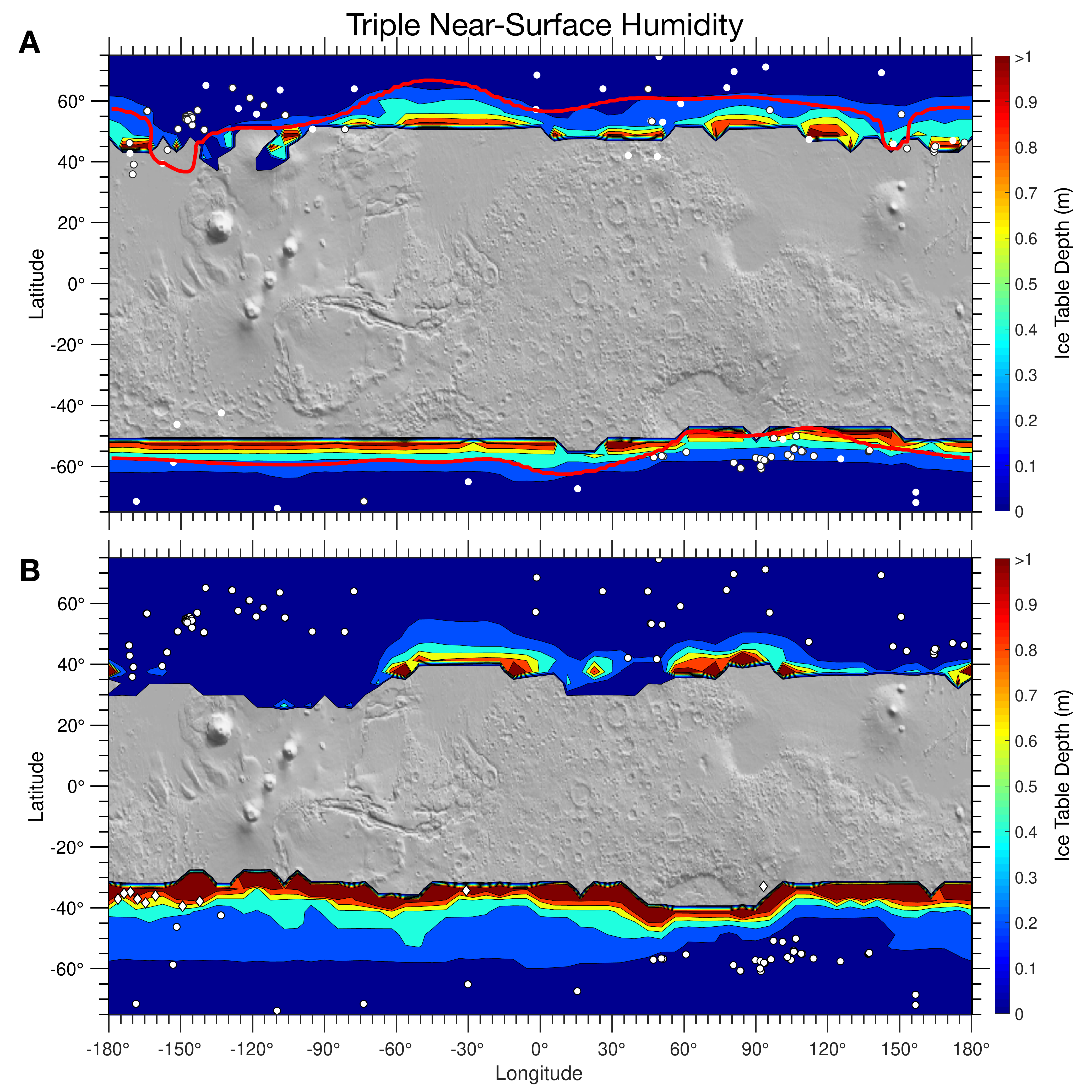}
\caption{Theoretical stability of subsurface water-ice with respect to diffusion for a) flat surfaces b) 30\textdegree~pole-facing slopes, using a triple near-surface humidity than those observed today. The red curve is the 10\% Water-Equivalent Hydrogen contour, which is a good proxy for the presence of water ice in the shallow subsurface \cite{PATHARE2018}. White dots indicate  exposed water ice along cliff scarps and impact craters as reported in \citeA{Dundas2021, Dundas2022} White diamonds indicate exposed water ice along gullies \cite{Khullericegullies}. 
    }
\end{figure}